\documentclass[fleqn,usenatbib]{mnras}

\usepackage[T1]{fontenc}
\usepackage{ae,aecompl}

\usepackage{graphicx,subfig}	
\usepackage{amsmath}	
\usepackage{amssymb}	

\title[Exploring the evolution of Reionisation using the Light Cone Effect]{Exploring the evolution of Reionisation using a wavelet transform and the light cone effect}

\author[C. M. Trott]{
Cathryn M. Trott$^{1,2}$\thanks{E-mail: cathryn.trott@curtin.edu.au}
\\
$^{1}$International Centre for Radio Astronomy Research, Curtin University, Bentley Australia\\
$^{2}$Australian Research Council Centre of Excellence for All-Sky Astrophysics (CAASTRO), Australia
}

\date{Accepted XXX. Received YYY; in original form ZZZ}

\pubyear{2016}

\begin{document}
\label{firstpage}
\pagerange{\pageref{firstpage}--\pageref{lastpage}}
\maketitle

\begin{abstract}
The Cosmic Dawn and Epoch of Reionisation, during which collapsed structures produce the first ionising photons and proceed to reionise the intergalactic medium, span a large range in redshift ($z\sim30-6$) and time ($t_{\rm age}\sim0.1-1.0$~Gyr). Exploration of these epochs using the redshifted 21~cm emission line from neutral hydrogen is currently limited to statistical detection and estimation metrics (e.g., the power spectrum) due to the weakness of the signal. Brightness temperature fluctuations in the line-of-sight (LOS) dimension are probed by observing the emission line at different frequencies, and their structure is used as a primary discriminant between the cosmological signal and contaminating foreground extragalactic and Galactic continuum emission. Evolution of the signal over the observing bandwidth leads to the `line cone effect' whereby the HI structures at the start and end of the observing band are not statistically consistent, yielding a biased estimate of the signal power, and potential reduction in signal detectability. We implement a wavelet transform to wide bandwidth radio interferometry experiments to probe the local statistical properties of the signal. We show that use of the wavelet transform yields estimates with improved estimation performance, compared with the standard Fourier Transform over a fixed bandwidth. With the suite of current and future large bandwidth reionisation experiments, such as with the 300~MHz instantaneous bandwidth of the Square Kilometre Array, a transform that retains local information will be important.
\end{abstract}




\section{Introduction}
The 21~cm signal evolves with redshift smoothly \citep{pritchard08,mesinger11}. Broadly, the evolution is a combination of amplitude (hydrogen spin temperature relative to the CMB temperature), and ionised fraction (the total number of neutral atoms decreases with time) components, which both vary spatially and temporally (ionised regions grow over time), \citep{furlanetto16,furlanetto06}. The combination interacts such that the signal evolves differently at different spatial scales, and the evolution is not necessarily monotonic with frequency. This leads to redshift ranges where some range of spatial modes evolve rapidly and non-monotonically \citep[see][for example evolution with redshift for a simulated signal]{datta12}.

The optimal bandwidth over which a line-of-sight transform should be performed, in order to balance signal bias due to evolution within the box, with loss of sensitivity from observing fewer data, was discussed by \citet{datta12}. However, this work performed only a cursory exploration of this balance by defining a simple metric for signal accuracy and not considering noise implications. Further, they did not consider the impact of foreground contamination on data of different bandwidths. Ultimately, information can be lost and the output results biased by performing a Fourier-like transform over a fixed bandwidth, and the degradation is spatial scale-dependent. This is due to the Fourier Transform using the full range of input channels, and treating them as statistically-equivalent in its computation of the spectral structure.

In this work, we aim to alleviate this deficiency of the Fourier Transform by considering a generalised transform that retains the key sinusoidal basis features of the Fourier Transform, but computes a local transform with variable scale. In doing so, we aim to formulate a hybrid transform to explore the EoR, and demonstrate the increased precision compared with standard techniques. The paper is outlined as follows: we introduce the expected EoR signal characteristics, and then describe how these are studied through spectral techniques. The Fourier Transform methodology for performing this is then reviewed, before the Morlet Transform is introduced and shown how it relates to the Fourier Transform. We then demonstrate its effect on 21~cm data, by studying and comparing both transforms in how they (1) transform foregrounds, and (2) transform the signal (through two toy models). We combine these with simulated noise, to compare the precision with which an optimal estimator could estimate cosmological parameters from the same underlying data, and with the Fourier and Morlet Transforms. This final step demonstrates the potential benefits of wavelet transforms for EoR 21~cm science.

Throughout we use a $\Lambda$CDM cosmology, with WMAP9 cosmological parameters \citep{hinshaw13}, $\Omega_Mh^2=0.14$, $\Omega_\Lambda=0.72$, and $H_0=100h$~km~s$^{-1}$~Mpc$^{-1}$.

\section{Methods}
\subsection{Signal structure}
The signal received by a radio telescope is the integrated contribution of source emission from the early Universe, extragalactic sources and Galactic emission, viewed through plasma refractive effects of the ionosphere and interplanetary medium. Extracting the signal of interest from the early Universe requires knowledge of the other sources of contamination. In general, our understanding of the contamination and propagation effects is incomplete, and we require additional techniques to perform the discrimination task.

Foreground contamination is the major systematic component to signal extraction, with the amplitude of extragalactic point sources (AGN and star-forming galaxies), Galactic synchrotron and Galactic free-free emission $\sim$10$^4$ times larger than the cosmological signal \citep{jelic08,jelic10,liu11,vedantham12,trott12}. Advantageously, these foreground sources emit continuum light, allowing a clear structural distinction from the emission-line brightness temperature fluctuations of the 21~cm signal, attributable to physical HI structures along the line-of-sight. The line-of-sight observation of HI is accessed through observations at different frequencies, where different spectral channels in the data map directly to different depths (redshifts) of HI emission (ignoring redshift-space distortions due to peculiar velocities). EoR experiments therefore sample a range of line-of-sight depths by observing the signal over a specific bandwidth \citep[e.g., ][]{parsons10,ali15,morales04,parsons14,dillon13,liu11,trott16}. Foreground treatments aim to utilise this different line-of-sight structure to discriminate contamination from signal. Treatments can typically be classed as `foreground avoidance' and `foreground suppression'.

Foreground avoidance aims to contain the foreground signal in a region of parameter space, and neglect this contamination in the foregoing analysis. In the statistical power spectrum experiment, a cube of observed signal (two angular and one line-of-sight direction) is Fourier Transformed along all three axes to obtain the signal as a function of spatial scale. The foreground contamination is smooth in frequency, naively containing it within the first few line-of-sight wavemodes. In reality, the chromaticity of the interferometer as an observing instrument changes the expected response such that smooth-spectrum foregrounds occupy a wedge-like region in $k_\bot-k_\parallel$ parameter space \citep{trott12,thyagarajan13,vedantham12}. Foreground avoidance is practised within the MWA and PAPER EoR experiments \citep{jacobs15,parsons10,beardsley13}. Foreground suppression aims to model the contamination and include it in the signal estimator. It can be further divided into non-parametric and parametric approaches, whereby the former uses knowledge of the low-frequency sky to motivate data-driven models of the actual sky signal \citep{liu11,dillon15,trott16}, while the latter fits a blind smooth signal to the spectral data with different degrees of sophistication \citep{bowman09,chapman12}. In both cases, the two-dimensional (2D) power spectrum, whereby the angular and line-of-sight scales are kept separate, is the initial data product, which is then further averaged spherically to the one-dimensional (1D) power spectrum using the relevant foreground treatment. The key point is that a Fourier-like transform is applied in the spectral dimension to capture the spatial structure of these modes. For a signal that evolves over the observation box, the signal is not statistically identical across the experimental bandwidth, leading to potential biases and imprecision in the signal extraction. This is the light cone effect \citep{datta12,ghara15}.

\begin{figure}
{
\includegraphics[width=.45\textwidth]{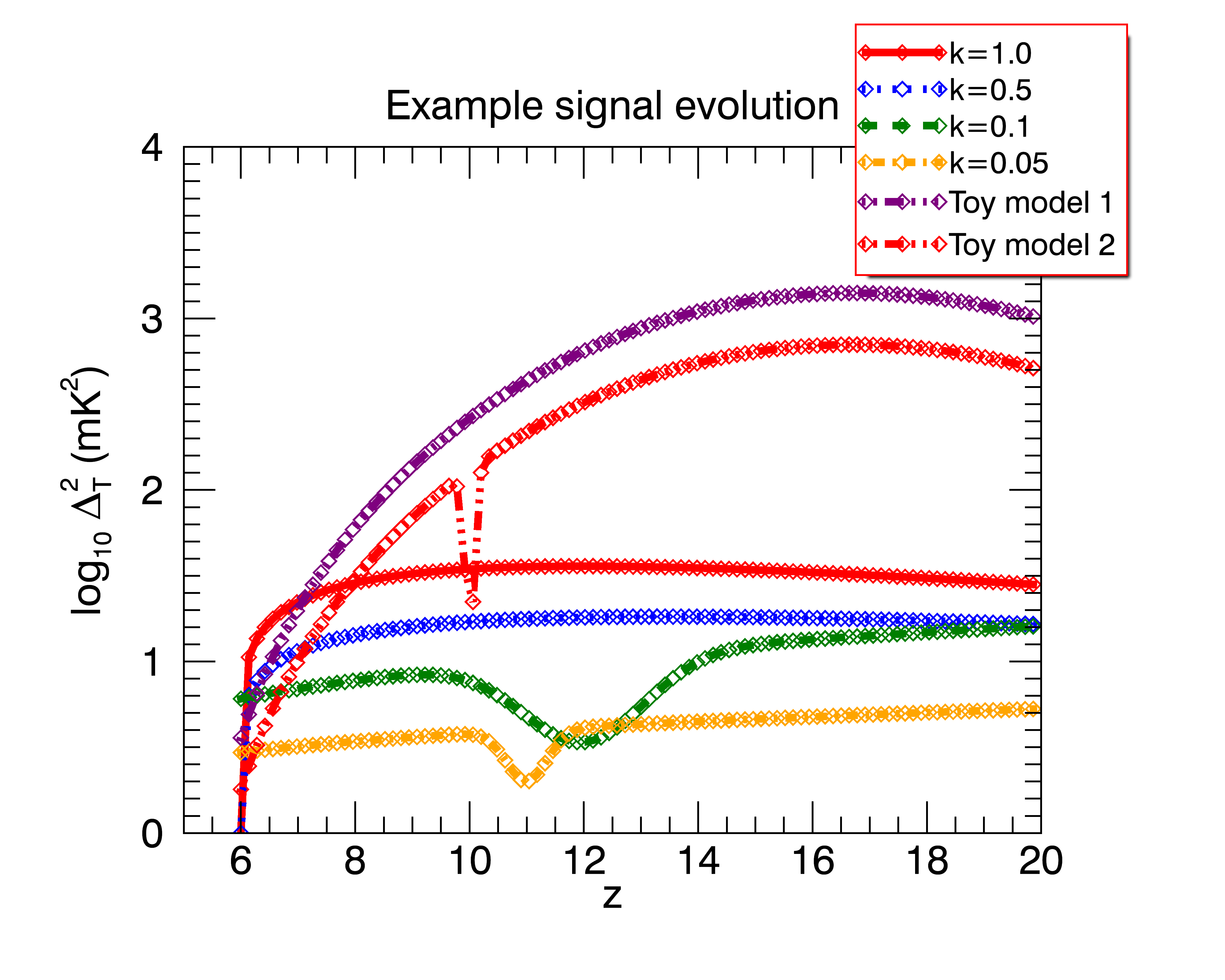}
}
\caption{Example evolution of the signal power as a function of redshift, for different spatial modes. Also shown is the signal for the two toy models used later ($k=0.01h$~Mpc$^{-1}$).}
\label{fig:model}
\end{figure}

The 21~cm signal evolution is model-dependent, but all models produce brightness temperature power as a function of redshift and spatial scale, that (1) can be non-monotonic, and (2) is functionally different between models of reionisation \citep[e.g.,][]{datta12}. Figure \ref{fig:model} shows mock dimensionless signal power curves as a function of redshift for a given angular scale.
It is clear that for the larger angular scales (small $k$), the existence of a trough at $z\sim$10--12 would lead to potential signal biases, because averaging over the bandwidth would destroy the curvature information. In general, in regions where $\frac{d^2\Delta^2}{dz^2}>0$, there will be power suppression relative to the observation box ends, whereas negative curvature would lead to power enhancement.

\subsection{The Fourier Power Spectrum}
The primary statistical measure of EoR/CD science is the spatial power spectrum (power spectral density), which quantifies the brightness temperature fluctuation power contained in a given spatial scale within an observation volume:
\begin{equation}
P(k) =  \frac{1}{V} \langle\tilde{T}(\vec{k})\tilde{T}^\ast(\vec{k})\rangle,
\end{equation}
where $T(\vec{k})$ is the temperature fluctuation (relative to the mean) at wavemode vector position $\vec{k}$, $\langle\rangle$ denotes an average over scales of that amplitude in the volume, $V$, and $k=\sqrt{u^2+v^2+\eta^2}$ is the scalar wavenumber.
The power spectrum measures the spatial covariance of a signal, integrated over a spatial volume, and corresponds to the Fourier Transform of the two-point correlation function (i.e., the autocorrelation function). Integration of the signal over the observation volume to form the power increases signal detectability because a large amount of data is summed to yield the output. For current generation EoR experiments, which are sensitivity-limited (e.g., Murchison Widefield Array (MWA){\footnote[1]{http://www.mwatelescope.org}} \citep{lonsdale09,tingay13_mwasystem}; Precision Array for Probing the Epoch of Reionization (PAPER){\footnote[2]{http://eor.berkeley.edu}} \citep{parsons10}; the Low Frequency Array (LOFAR){\footnote[3]{http://www.lofar.org}} \citep{vanhaarlem13}; the Long Wavelength Array (LWA){\footnote[4]{http://lwa.unm.edu}} \citep{ellingson09}; and Hydrogen Epoch of Reionization Array (HERA){\footnote[5]{http://reionization.org}}), a statistical metric is required for signal detectability.

The power spectrum of temperature fluctuations can be computed from an image cube, via a three-dimensional Fourier Transform. It can also use native interferometer visibilities, which measure the spatial coherence of signals from the sky, and implicitly perform an angular, two-dimensional Fourier Transform of the sky signal. A direct power spectrum from visibilities has been used with current EoR experiments \citep{trott16,choudhuri14}. The final Fourier Transform, along the line-of-sight, transforms frequency channels to $\eta$ modes over a fixed bandwidth, $B=N_{\rm chan}\Delta\nu$:
\begin{eqnarray}
\tilde{T}(u,v,\eta) &=& \mathcal{F}\tilde{T}(u,v,\nu)\\
&=& \Delta\nu \displaystyle\sum_{\nu_i=0}^{N_{\rm chan}-1} \tilde{T}(u,v,\nu_i) \exp{(-2\pi{i}\nu_i\eta)}.
\end{eqnarray}
Performing the transform over a large bandwidth leads to more robust characterisation of the spatial structure and lower thermal noise, but increases signal bias by incorporating information that is not statistically consistent (hydrogen brightness temperature distribution that evolves through the volume). It is this balance that we wish to treat more formally with the wavelet transform.

\subsection{The Morlet Transform}
\begin{figure}
{
\includegraphics[width=.45\textwidth]{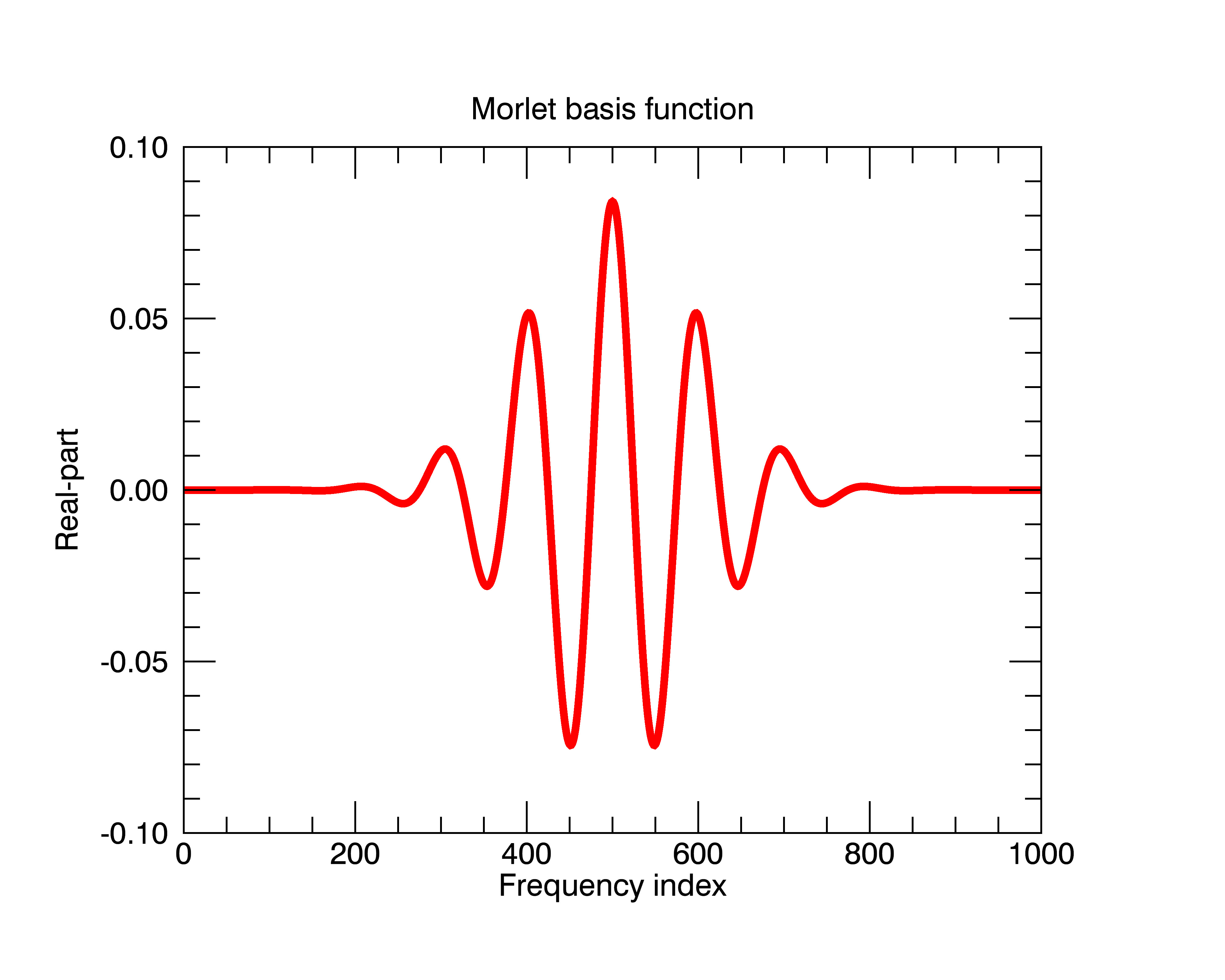}
}
\caption{Example real-part of a Morlet basis function, showing a Gaussian-localised Fourier response translated to a central frequency index of 500.}
\label{fig:morlet_example}
\end{figure}

The Morlet Transform is a Continuous Wave Transform \citep{goupillaud84}. It is characterised by a Fourier basis, enclosed within a Gaussian envelope, and is designed to provide a balance between resolution in the two dual-Fourier parameters (here, observation frequency and line-of-sight spatial scale, $\eta\propto{k_{\parallel}}$). It is widely used in other fields of science to extract local spectral information from the data. In the context of EoR, it allows computation of the power in spatial scales, local to a given redshift, thereby balancing signal evolution with optimal bandwidth. Note that the Morlet is not the only transform appropriate for this task; in general, any transform that provides a well-behaved, localised envelope to a Fourier basis, and maps to physically-relevant parameters (e.g. the line-of-sight wavenumber $\eta$) would be suitable. Figure \ref{fig:morlet_example} shows the real-part of an example complex-valued Morlet basis function, highlighting the key features of the Fourier carrier and the localising Gaussian envelope.

The Morlet Wavelet uses the Morlet Transform basis functions to describe the local, spectral properties of a signal. We define the scaled and translated Morlet Wavelet Transform (MWT) basis function, $\Psi$, by:
\begin{equation}
\Psi(\nu_i;\eta,\nu_c) = \exp{\left(-\eta^2(\nu_i-\nu_c)^2 \right)}\exp{\left( 2\pi{i}\eta(\nu_i-\nu_c) \right)},
\end{equation}
where $\nu$ are the observed frequency channels (MHz), $\eta$ is the line-of-sight spatial scale (Fourier dual, MHz$^{-1}$), and $\nu_c$ is the observation frequency associated with a given redshift. The discrete MWT of signal $X(\nu)$ is therefore:
\begin{eqnarray}
\Psi(\eta,\nu_c) &=& \frac{\sqrt{|\eta|}\Delta\nu}{\pi^{1/4}}\times\\\nonumber
&&\displaystyle\sum_{i=0}^{N-1} X(\nu_i) \exp{\left(-\eta^2(\nu_i-\nu_c)^2 \right)}\exp{\left( 2\pi{i}\eta(\nu_i-\nu_c) \right)},
\end{eqnarray}
where the constant scale normalises the energy contained within the wavelet, and $\Delta\nu$ is the signal frequency resolution. For $N$ spectral channels, where the total bandwidth is BW$=N\Delta\nu$, the transform is computed discretely at:
\begin{eqnarray}
\nu_c &=& [0,\Delta\nu,2\Delta\nu,\dots,(N-1)\Delta\nu]\\
\eta &=& \left[0,\frac{1}{N\Delta\nu},\frac{2}{N\Delta\nu},\dots,\frac{N-1}{2(N\Delta\nu)}\right],
\end{eqnarray}
to align with the measured frequencies of the Fourier Transform. This leads to correlations between neighbouring parameters in the 2D parameter space, which must be accounted for in the analysis.

The data are measured brightness temperature fluctuations at a given angular scale, $T(\nu;u,v)$~mK, where $u,v$ are angular Fourier modes measured by an interferometer, and $k_\bot\propto\sqrt{u^2+v^2}$. Therefore, each 1D spectral data series maps to a 2D Morlet parameter space, yielding a set of Morlet power spectra (one for each angular scale, $k_\bot$).

Due to the substantial evolutionary history encompassed by a wide bandwidth experiment, the conversion from line-of-sight scale measured in inverse MHz and the comoving cosmological frame (measured in inverse Mpc), warps the square $\nu_c-\eta$ parameter space into a non-square $z-k_\parallel$ space, according to:
\begin{eqnarray}
z(\nu_c) &=& \frac{f_{21}}{\nu_c}-1\\
k_\parallel(\eta,z) &=& \eta\frac{2\pi {H}_0 f_{21}\sqrt{\Omega_M(1+z)^3+\Omega_\Lambda}}{c(1+z)^2},
\end{eqnarray}
where $f_{21}\approx{1420}$~MHz, $c$ is the speed of light, $H_0$ is the Hubble constant and $\Omega_M$ and $\Omega_\Lambda$ are the matter and dark energy density of the Universe.

\begin{figure}
{
\includegraphics[width=.45\textwidth]{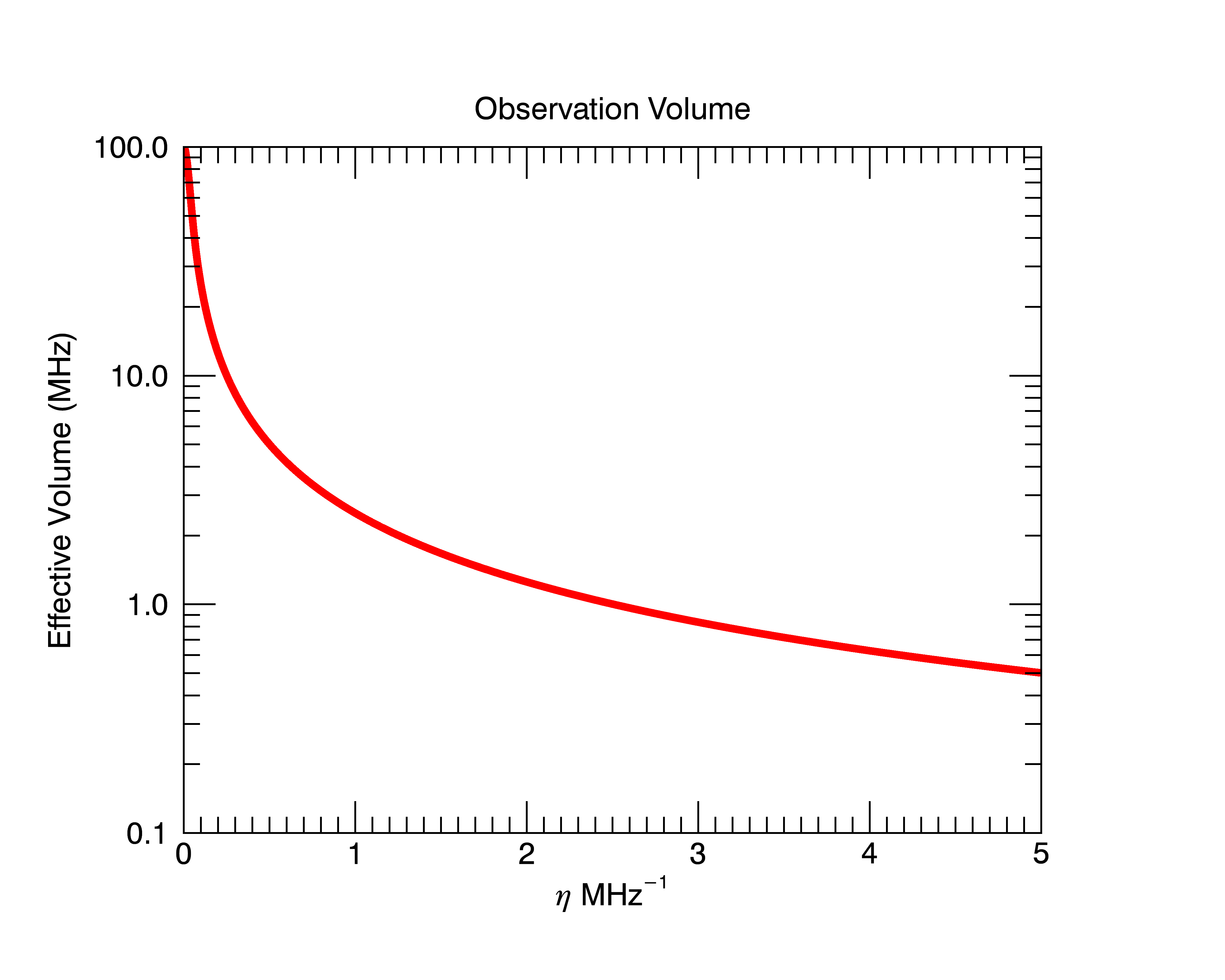}
}
\caption{Effective observation depth (bandwidth) as a function of line-of-sight scale, $\eta$ for a 100~MHz dataset transformed through a Morlet wavelet, with a central frequency of 150~MHz, $z$~=~8.5.}
\label{fig:obs_vol}
\end{figure}

Finally, as with the Fourier Transform power spectrum, the Morlet power spectrum (MPS) is computed by taking the absolute square of the mode signals and normalising by the observation bandwidth. For a Gaussian envelope containing the ``observation'' depth, the normalisation is given by the weighted integral over the envelope (except for small $\eta$ values where the Gaussian window is truncated),
\begin{eqnarray}
B(\eta) &=& \frac{\sqrt{2\pi}}{\eta}\,\,{\rm MHz},\\
B(k_\parallel) &=& \frac{(2\pi)^{3/2} {H}_0 f_{21}\sqrt{\Omega_M(1+z)^3+\Omega_\Lambda}}{k_\parallel{c}(1+z)^2}\,\,{\rm Mpc}.
\end{eqnarray}
Similarly, the field-of-view of a fixed-size aperture with effective area, $A_{\rm eff}$, is frequency-dependent. As a function of redshift, the effective observation volume ($h^{-3}$~Mpc$^3$) is:
\begin{equation}
V(k_\parallel,z) = B(k_\parallel){\rm DM}(z)^2\frac{c^2(1+z)^2}{f21^2A_{\rm eff}} \,\,h^{-3}{\rm Mpc^3},
\end{equation}
giving a final power spectrum,
\begin{equation}
P_{\rm MWT}(z,k_\parallel;k_\bot) = \frac{1}{V(k_\parallel,z)}|\Psi(k_\parallel,\nu_c)|^2\,\,{\rm mK^2}h^{-3}{\rm Mpc^3}.
\end{equation}
Figure \ref{fig:obs_vol} plots the effective observation bandwidth as a function of scale, $\eta$, for a 100~MHz bandwidth dataset.
Thus, although the total observational bandwidth used in this work is 100~MHz, each line-of-sight scale has a smaller effective bandwidth, with the smallest scales transformed over bandwidths of $<$2~MHz.

\begin{figure*}
{
\includegraphics[width=.98\textwidth]{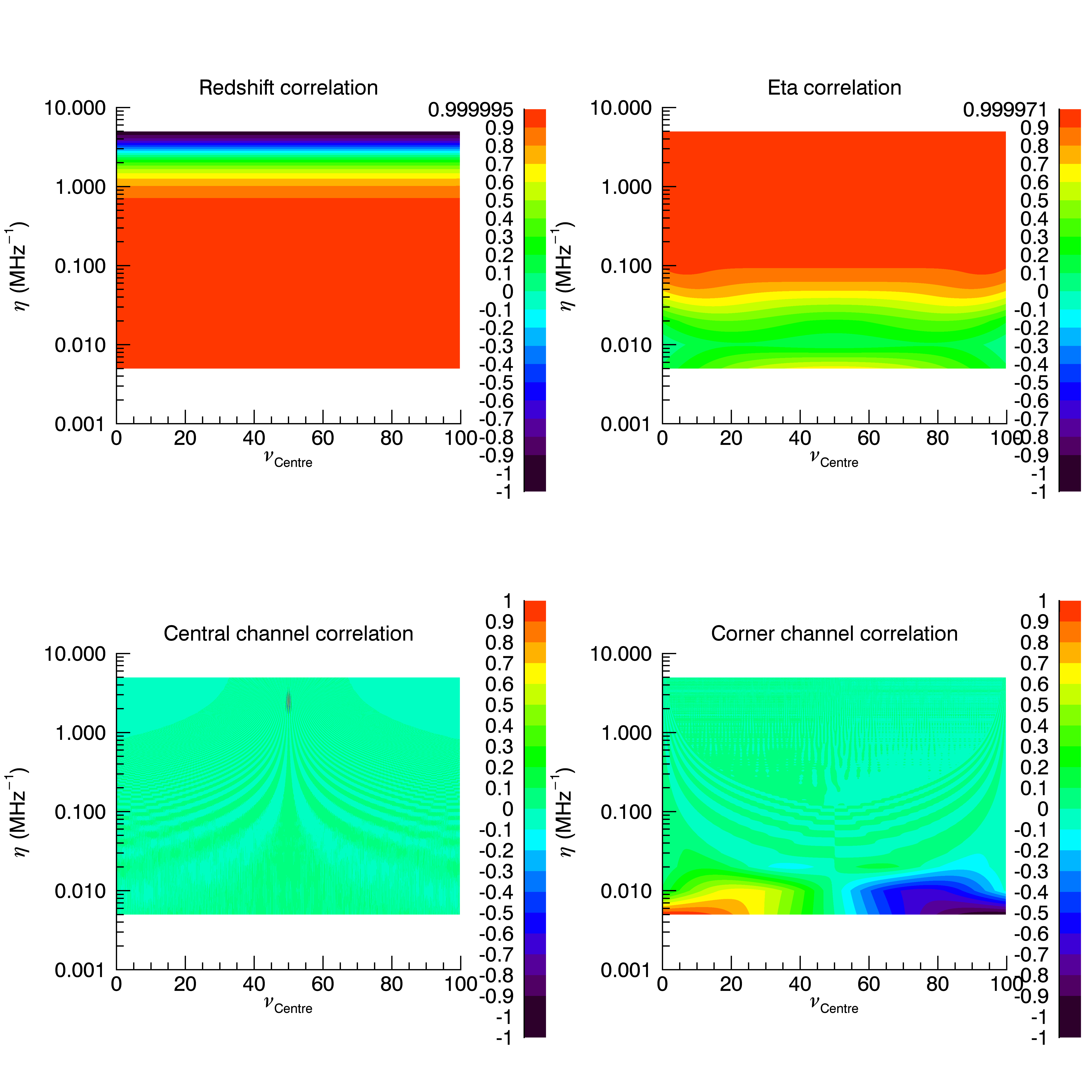}
}
\caption{The correlation coefficient, $\xi$, between adjacent cells in the redshift (top-left) and $\eta$ (top-right) dimensions. The correlation of a single cell in the centre (bottom-left) and corner (bottom-right) of the parameter space with all other cells is also shown.}
\label{fig:correlation}
\end{figure*}

At high spatial frequencies (large $\eta$ values), the signal is highly-localised and the effective bandwidth is small. This leads to substantial correlation between neighbouring locations in parameter space. Further, the sliding translation of the central observing redshift leads to correlation in that dimension. Figure \ref{fig:correlation} displays the correlation coefficient, $\xi$, between adjacent cells in the redshift (top-left) and $\eta$ (top-right) dimensions, quantified by computing the overlap in basis functions, where the coefficient for correlation in the $\eta$ dimension is given by:
\begin{equation}
\xi(\eta_i,\nu_c) = \frac{\langle \Psi(\eta_i,\nu_c)^\dagger\Psi(\eta_{i+1},\nu_c) \rangle}{ \sqrt{\langle\Psi(\eta_i,\nu_c)^\dagger\Psi(\eta_i,\nu_c)\rangle\langle\Psi(\eta_{i+1},\nu_c)^\dagger\Psi(\eta_{i+1},\nu_c)\rangle}},
\end{equation}
and the $\langle\rangle$ denote an average over frequency. The correlation of a single cell in the centre (bottom-left) and corner (bottom-right) of the parameter space with all other cells is also shown. In both cases, the correlation length is short.

\subsection{Foregrounds viewed through transforms}\label{sec:fg}
In previous work, the statistical signature in the $(k,\nu)$ space of spectrally-smooth, Poisson-distributed extragalactic point sources has been described \citep{trott12,trott16,liu11,hazelton13,thyagarajan15a,thyagarajan15b}. We use the same formalism as described in \citet{trott16} to motivate the signature of foregrounds through the Morlet Wavelet Power Spectrum. We reproduce the key components of the formalism here, and refer the reader to the previous work for more depth.

\begin{figure*}
{
\includegraphics[width=.98\textwidth]{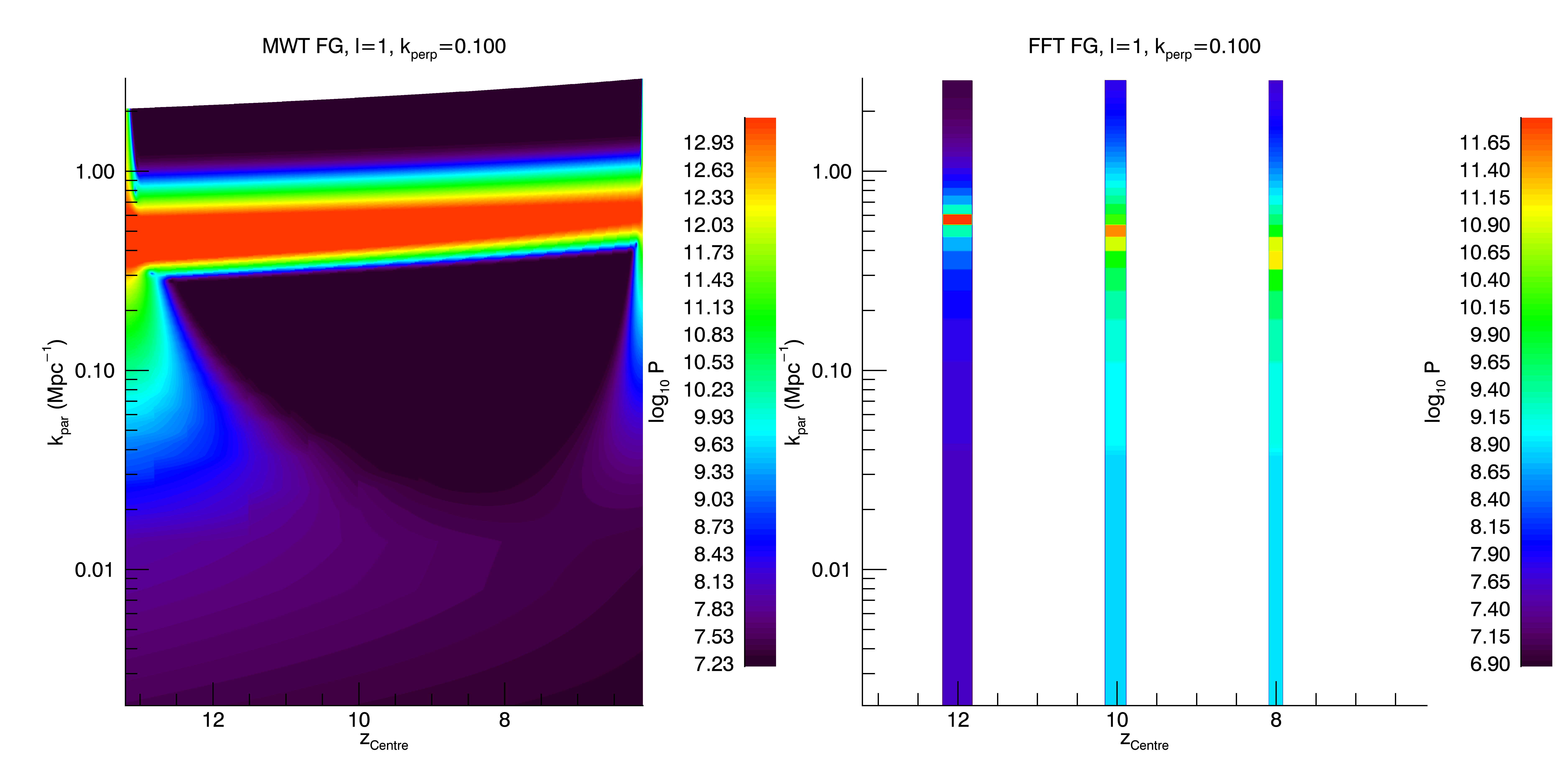}
}
\caption{Power from Morlet (left) and Fourier (right) Transforms for a single point source located at the horizon ($l=1$). The Fourier Transform is performed at three specific redshifts, with a fixed bandwidth of 8~MHz, while the Morlet Transform is performed across all sampled data in the 100~MHz observational band.}
\label{fig:fgsingle}
\end{figure*}

Initially, we can demonstrate the differing outputs of the Fourier and Morlet Transforms for a simple foreground by considering a single smooth spectrum source across frequency. The interferometer sampling function is chromatic, leading to a wrapping of phase as a function of frequency, which is linearly-dependent on the baseline length and the position of the source in the sky relative to the phase centre. This is the origin of the wedge structure in the 2D power spectrum. We apply this phase wrap, and compute the power spectrum for a single source located at the horizon. Figure \ref{fig:fgsingle} displays the output of the two transforms.
The Fourier Transform is performed at three specific redshifts with a fixed bandwidth of 8~MHz, while the Morlet Transform yields power at all redshifts in the data using the full 100~MHz observation band. Both show the peak associated with a source delayed relative to the phase centre, but with a different gradient with redshift. The reduced transverse comoving distance for lower redshifts transforms to a lower value of $k_\parallel$ for the peak. However, the Morlet Transform is performed over a much larger fractional bandwidth, leading to faster wrapping of the phase for lower redshifts (higher frequency). This works against the comoving distance transform, and yields the opposite gradient. Nonetheless, the structures are similar between the two transforms.

We can now extend the analysis to consider all sources in the sky, and their relative positions and flux densities. This extends the model to a statistical description of the source distribution. The aim is to take a statistical description of foregrounds in the sky image space ($l,m,\nu$), and transform this to angular Fourier space ($u,v,\nu$) and then wavemode space ($u,v,\eta$), while incorporating and retaining the instrumental response to the signal (i.e., frequency-dependent primary beam shape, chromatic interferometer sampling function). The point source foregrounds follow a power-law number density as a function of flux density, and are Poisson-distributed in the sky. The sources are expected to be uncorrelated between sky locations, and the transformation to angular Fourier space assumes that the individual sky location covariance matrices can be added. The covariance between frequency channels $\nu^\prime$ and $\nu^{\prime\prime}$ at angular mode, $k=\sqrt{u^2+v^2}$ wavelengths for a circularly-symmetric beam is given by a Hankel transform:
\begin{eqnarray}
&&\boldsymbol{C}(\nu^\prime,\nu^{\prime\prime};k) =\frac{\alpha}{3-\beta} \left( \frac{\sqrt{\nu^{\prime\prime}\nu^\prime}}{\nu_{\rm low}} \right)^{-\gamma} \frac{S_{\rm max}^{3-\beta}}{S_0^{-\beta}}\\\nonumber
 &\times& \int_0^\infty B(l;\nu^{\prime\prime})B(l;\nu^\prime) J_0\left({2\pi(kl)(\nu^{\prime\prime}-\nu^\prime)}\right) ldl \, {\rm Jy^2},
\end{eqnarray}
where $S_{\rm max}$, $\alpha$ and $\beta$ are all parameters of number counts model, $l=|\vec{l}|$ is the direction cosine in the sky, $B(\nu)$ is the primary beam as a function of frequency, and $\nu_{\rm low}$ is the lowest observation frequency. This model incorporates a frequency-dependent beam and the interferometer chromatic sampling that leads to the wedge feature, with smaller angular scales (large $k$) having the most foreground leakage to higher line-of-sight spatial modes. In this work, we consider a 10~m diameter station, with a corresponding field-of-view at 150~MHz of $\sim$~11~degrees.

\begin{figure*}
{
\includegraphics[width=.98\textwidth]{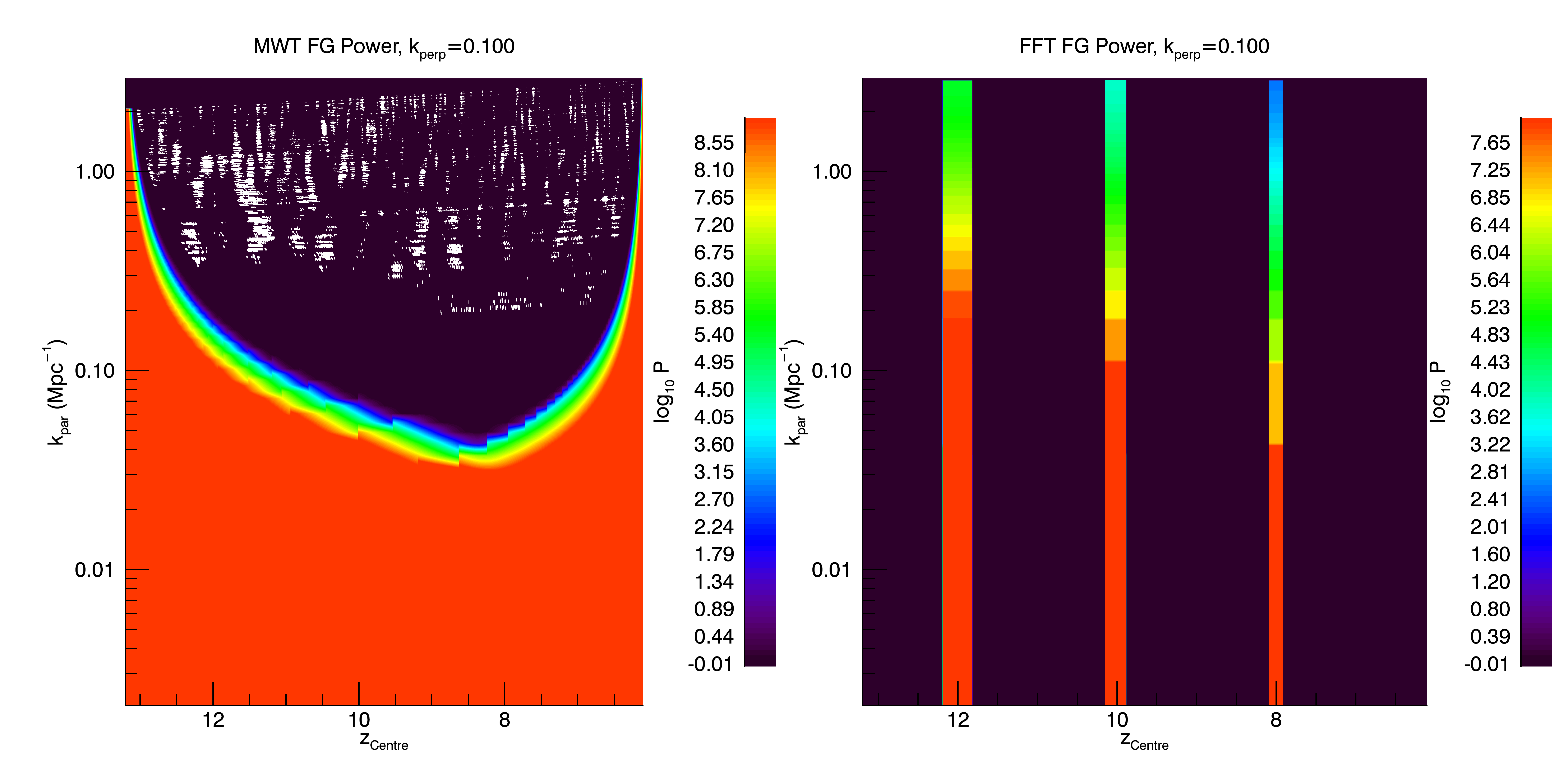}
}
\caption{Power from Morlet (left) and Fourier (right) Transforms for a statistical point source model. The Fourier Transform is performed at three specific redshifts, with a fixed bandwidth of 8~MHz, while the Morlet Transform is performed across all sampled data in the 100~MHz observational band.}
\label{fig:fgcov}
\end{figure*}

The transform of the covariance matrices through a line-of-sight Fourier ($\mathcal{F}$) or Morlet transform ($\mathcal{M}$) is:
\begin{eqnarray}
&&\boldsymbol{C}_{\rm FPS}(\eta^\prime,\eta^{\prime\prime};k_0) = \mathcal{F}^\dagger \boldsymbol{C}(\nu^\prime,\nu^{\prime\prime};k) \mathcal{F}\\
&&\boldsymbol{C}_{\rm MPS}(\eta^\prime,\eta^{\prime\prime};k_0) = \mathcal{M}^\dagger \boldsymbol{C}(\nu^\prime,\nu^{\prime\prime};k) \mathcal{M}.
\end{eqnarray}
The Morlet Transform yields the foreground signature at all redshifts for the given angular scale, and the Fourier Transform is again sampled at three redshifts, each with 8~MHz bandwidth. Figure \ref{fig:fgcov} displays the power from this statistical foreground model.
Note that in these figures, there is no power from the horizon sources, because they are highly attenuated by the primary beam. Furthermore, the shape of the Morlet Transform is qualitatively different from the horizon-source model, because the primary beam field-of-view reduces for lower redshift. This therefore more accurately reflects the foreground contribution to the power in the Morlet Transform.

The larger available instantaneous bandwidth of the MWT yields better foreground containment in low $k_\parallel$ modes, compared with the FFT over a smaller bandwidth, simply due to the reduced spectral leakage of having a larger number of sampled frequencies. This key feature of the wavelet transform is crucial for current and future 21~cm experiments, where foreground contamination is found to be, and expected to remain, a source of primary systematic error in the measurements.

\subsection{Toy models}\label{sec:toy}
We construct two toy models to describe signal evolution as a function of angular scale. The temperature distribution along a sight-line is modelled as a Gaussian random field, with amplitude and correlation length characterising the evolution of signal amplitude and bubble size. The key features include the reduction of brightness temperature with time (relative to the CMB), and increase in HII bubble size with time. The second model includes a localised trough in the brightness temperature evolution. These models are not intended to provide a physical model of an actual reionisation history, but rather retain some of the key evolutionary features in an analytic form.

For both models, the base brightness temperature distribution on the sky, as a function of redshift, is modelled as:
\begin{equation}
T_B(l,m,\nu) \sim \mathcal{N}(0,A(z)^2) \ast \exp{(-R^2/\sigma(z)^2)},
\end{equation}
where $\mathcal{N}$ indicates a Gaussian-distributed variable with mean zero and variance, $A(z)^2$, `$\ast$' is the convolution operator, and $R$ is the comoving scalar distance. This form describes Gaussian-distributed temperature fluctuations along the line-of-sight, which evolve with redshift (frequency) and are correlated in comoving length over a scale $\sigma(z)$. The evolution functions, $A$ and $\sigma$ are given by:
\begin{eqnarray}
A(z) &=& \frac{A_s}{z_s-z_e}(z-z_e),\\
\sigma(z) &=& \frac{\sigma_{\rm ref}}{z_s-z_{\rm ref}}(z_s-z),
\end{eqnarray}
where $A_s$ (mK) and $\sigma_{\rm ref}$ ($h^{-1}$~Mpc) are reference amplitudes and sizes, and `$s$', `$e$', `${\rm ref}$' are the reionisation start, end and reference redshifts and scales. We Fourier transform this in the angular modes ($(l,m)-(u,v)$) to obtain the brightness temperature distribution as sampled by the interferometer:
\begin{eqnarray}
&\tilde{T}_B&(u,v,\nu)\\\nonumber
&\sim& \mathcal{CN}(0,A(z)^2)\exp{(-|u|^2\sigma(z)^2)} \ast \exp{(-\nu^2/\sigma(z)^2)}\\
&\sim& \mathcal{CN}\left(0,\left(\frac{A_s}{(z_s-z_e)^2}(z-z_e)^2\right)^2\right)\label{eqn:tb}\\\nonumber
&&\times \exp{\left(-\frac{k_\bot^2\sigma_{\rm ref}^2}{(z_s-z_{\rm ref})^2}(z_s-z)^2\right)} \ast \exp{(-z^2/\sigma(z)^2)},\nonumber
\end{eqnarray}
which yields the ensemble-averaged brightness temperature fluctuation power,
\begin{eqnarray}
&&\Delta^2_{21}(u,v,\nu) = \left(\frac{A_s}{(z_s-z_e)^2}(z-z_e)^2\right)^2\\
&&\times\exp{\left(-2\frac{(u^2+v^2)\sigma_{\rm ref}^2}{(z_s-z_{\rm ref})^2}(z_s-z)^2\right)} \ast \exp{(-2z^2/\sigma(z)^2)},\nonumber
\end{eqnarray}
with units, mK$^2$~Mpc$^4$.
Equation \ref{eqn:tb} describes the input signal for both the Fourier and Morlet line-of-sight transforms for Toy Model 1. We randomly generate a large number of Gaussian-distributed fluctuations according to this expression, and use these to sample the final power spectrum. The average brightness temperature variance for this model is also shown in Figure \ref{fig:model}.

Toy Model 2 uses the same base model, but with a Gaussian-shaped absorption feature at redshift, $z_{\rm dip}$, aimed to mimic a more realistic signal, and one with localised structure. In this case, the brightness temperature distribution is given by:
\begin{eqnarray}
&&T_B(l,m,\nu) \sim \\
&&\mathcal{N}(0,A(z)^2) \ast \exp{(-R^2/\sigma(z)^2)(1.-\exp{(z-z_{\rm dip})^2/\Delta{z}^2})^2)},
\end{eqnarray}
where $\Delta{z}$ parametrises the breadth of the absorption feature. This model is shown as Toy Model 2 in Figure \ref{fig:model}.

Table \ref{table:params} describes the parameters for the model, which is centred on $\nu=$~150~MHz.
\begin{table}
	\centering
	\caption{Parameters of reionisation used in the toy model.}
	\label{table:params}
	\begin{tabular}{l|cc} 
		\hline
		Parameter & Value & Units\\
		\hline
		$z_s$ & 13.3 & \\
		$z_e$ & 5.0 & \\
		$z_{\rm ref}$ & 7.0 & \\
		$z_{\rm centre}$ & 8.5 & \\
		$\sigma_{\rm ref}$ & 100 & $h^{-1}$~Mpc \\
		$A_s$ & 27~$\sqrt{(1+z_s)/(11)}$ & mK\\
		$z_{\rm dip}$ & 10 & \\
		$\Delta{z}^2$ & 0.02 & \\
		\hline
	\end{tabular}
\end{table}

\subsection{Parameter estimation precision}
To demonstrate the utility of the broad bandwidth afforded by the Morlet Transform, we compute the precision with which an ideal estimation procedure could measure two key parameters of the toy model 21~cm signals, in the presense of foregrounds with unknown parameters, and thermal noise. We use the Cramer-Rao Bound \citep[CRB,][]{kay93,kay98,vantrees01} to compute the minimum variance on the set of unknown parameters, which is computed via the Fisher Information Matrix (FIM), describing the information content of the data with regard to a particular parameter.

For complex-valued data that are distributed as a generalised multivariate Gaussian, the elements of the FIM for parameters of the data covariance matrix, ${\boldsymbol{C(\theta)}}$ are given by:
\begin{equation}
\mathcal{I}_{ij} = {\rm tr}\left[{{\boldsymbol{C^{-1}(\theta)}}} \frac{\partial{{\boldsymbol{C}}}}{\partial{\theta_i}} {{\boldsymbol{C^{-1} (\theta)}}} \frac{\partial{{\boldsymbol{C}}}}{\partial{\theta_j}}\right],
\end{equation}
where `tr' denotes the matrix trace. We assume that the data are $(u,v,\nu)$ complex-valued gridded visibilities. The coherent averaging of many measured visibilities within a given $(u,v,\nu)$ cell yields data that have zero mean, and covariance matrix described by a sum of 21~cm signal, foreground signal and noise, and use the toy model and foreground point source model described above. For the thermal noise, we form a simple model based on a generic interferometer with uniform filling of the $uv$-plane, and a $T$ hour total experiment.

Specifically, the data covariance matrix is:
\begin{equation}
\boldsymbol{{C}} = \boldsymbol{{C}}_{\rm 21} + \boldsymbol{{C}}_{\rm FG} + \boldsymbol{{C}}_{\rm N},
\end{equation}
where each component, measured in mK$^2{h}^{-4}$~Mpc$^4$,
\begin{eqnarray}
\boldsymbol{{C}}_{\rm 21} &=& A(\nu_1)A(\nu_2)\exp{\left(-2\frac{(u^2+v^2)\sigma_{\rm ref}^2}{(z_s-z_{\rm ref})^2}(z_s-z_1)(z_s-z_2)\right)}\nonumber\\ &&\ast \exp{(-2(z_1-z_2)^2/\sigma(z)^2)}\frac{\lambda_1^2\lambda_2^2DM(z)^4}{A_{\rm eff}^2},\\
\boldsymbol{{C}}_{\rm FG} &=& \frac{\alpha}{3-\beta} \left( \frac{\sqrt{\nu_1\nu_2}}{\nu_{\rm low}} \right)^{-\gamma} \frac{S_{\rm max}^{3-\beta}}{S_0^{-\beta}} \frac{\pi{c^2}\epsilon^2}{D^2}\frac{1}{\nu_1^2+\nu_2^2} \nonumber\\ &&\times \exp{\left( \frac{-u^2c^2f_{\nu}^2\epsilon^2}{4(\nu_1^2 + \nu_2^2)D^2} \right)}\frac{A_{\rm eff}^2DM(z)^4}{(2k)^2},\\
\boldsymbol{{C}}_{\rm N} &=& \frac{4T_{\rm sys}^2}{A_{\rm eff}}\frac{1}{N_{\rm vis}\Delta\nu{T}\sqrt{N_k}}DM(z)^4,
\end{eqnarray}
where `1' and `2' denote the two frequencies/redshifts of the covariance, and $T_{\rm sys}=180,000\,{\rm mK}(\nu/180)^{-2.6}$ is the sky-dominated system temperature. Here we describe the signal covariance for Toy Model 1. The noise parameters, $N_{\rm vis}$ and $N_k$ denote the number of coherent visibilities within each cell, and incoherent visibilities summing in power at a given $k=\sqrt{u^2+v^2}$, respectively. $\boldsymbol{C^{-1} (\theta)}$ denotes the matrix inverse evaluated at the parameter values, $\boldsymbol{\theta}$.

We consider a model where we aim to estimate the amplitude of the brightness temperature fluctuations ($A_s$, mK), and the reference comoving bubble size ($\sigma_{\rm ref}$, $h^{-1}$~Mpc), at the reference redshift, in the presence of unknown foreground number counts power law slope ($\beta$) and amplitude ($\alpha$, Jy$^2$~sr$^{-1}$) (Toy Model 1). For Toy Model 2, we additionally estimate the redshift of the absorption feature, $z_{\rm dip}$. The unknown foregrounds are designed to mimic the reality of estimating a weak signal in the presence of uncertain bright contaminants. We use the values from Table \ref{table:params} and \ref{table:toymodel2}, and analytically differentiate with respect to the four (five) unknown parameters. The matrix inverse is performed computationally. The experiments undertaken are: (1) use of the full 100~MHz bandwidth (MWT dataset); (2) use of individual 8~MHz bandwidths across the band (FFT dataset); (3) combination of individual, contiguous 8~MHz bandwidths tiling the full 100~MHz band.

\section{Results}
\label{sec:results} 
\begin{figure*}
\subfloat[Power at angular mode $k_\bot$~=0.001~$h^{-3}$~Mpc$^3$.]{
\includegraphics[width=.45\textwidth]{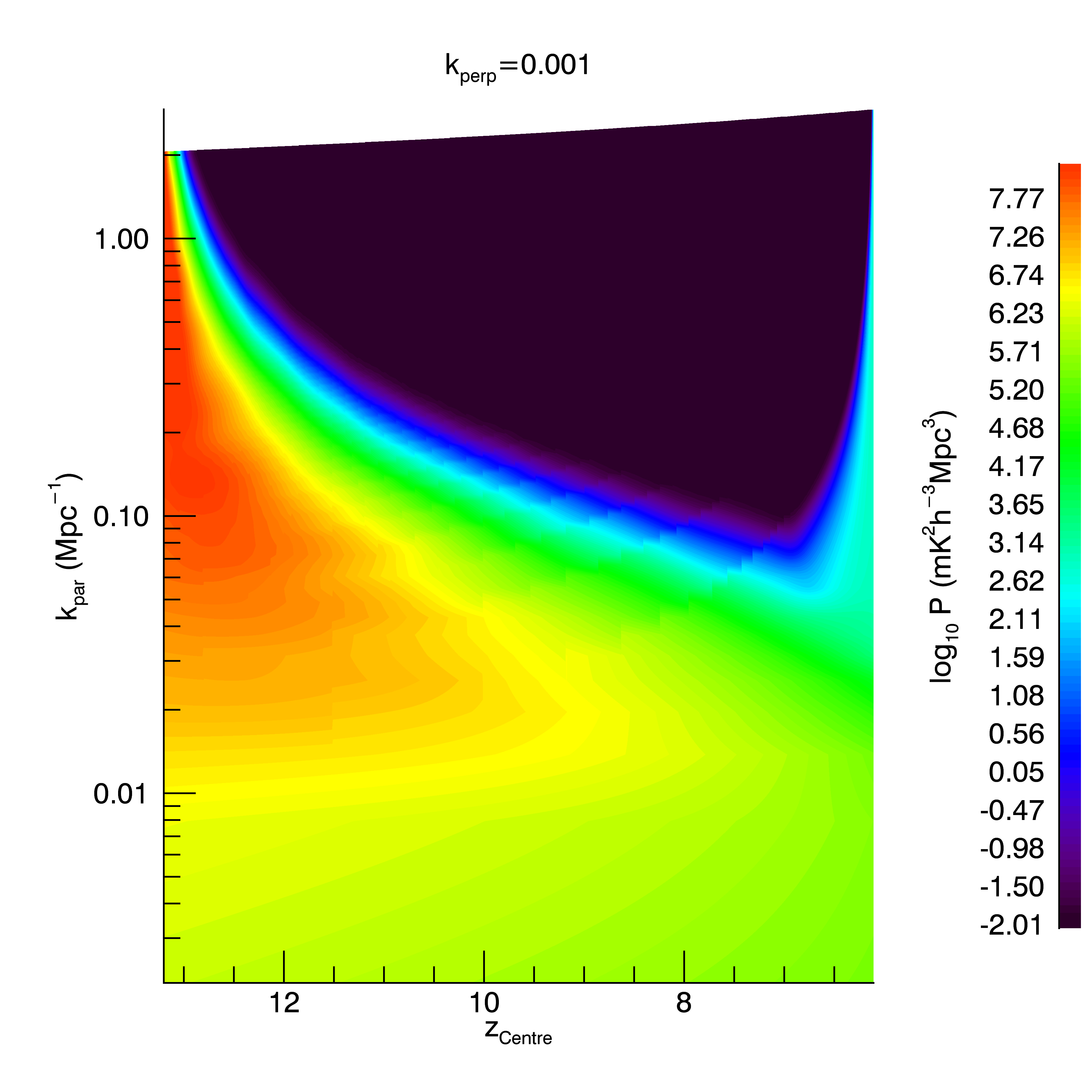}
}
\subfloat[Power at angular mode $k_\bot$~=0.010~$h^{-3}$~Mpc$^3$.]{
\includegraphics[width=.45\textwidth]{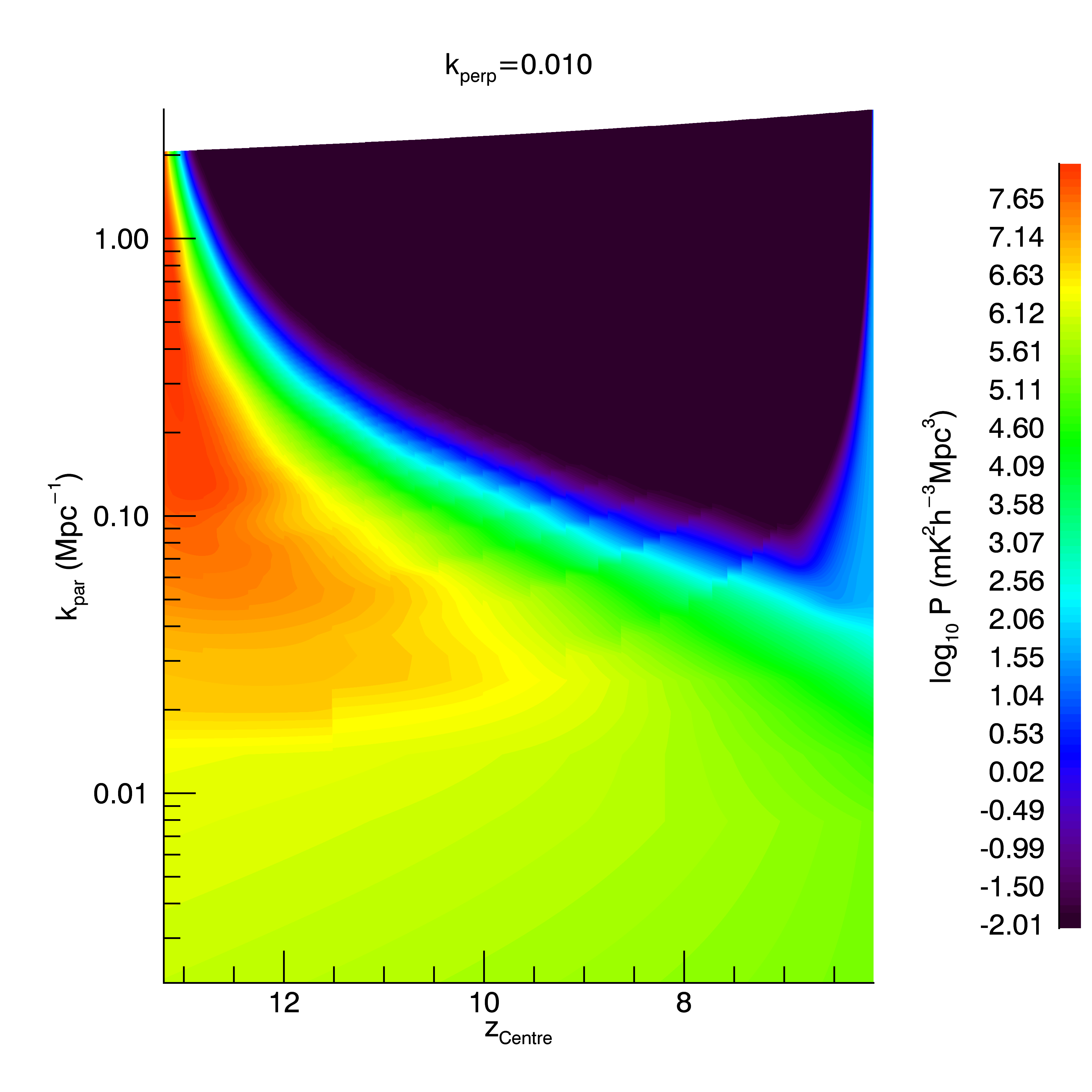}
}\\
\subfloat[Power at angular mode $k_\bot$~=0.100~$h^{-3}$~Mpc$^3$.]{
\includegraphics[width=.45\textwidth]{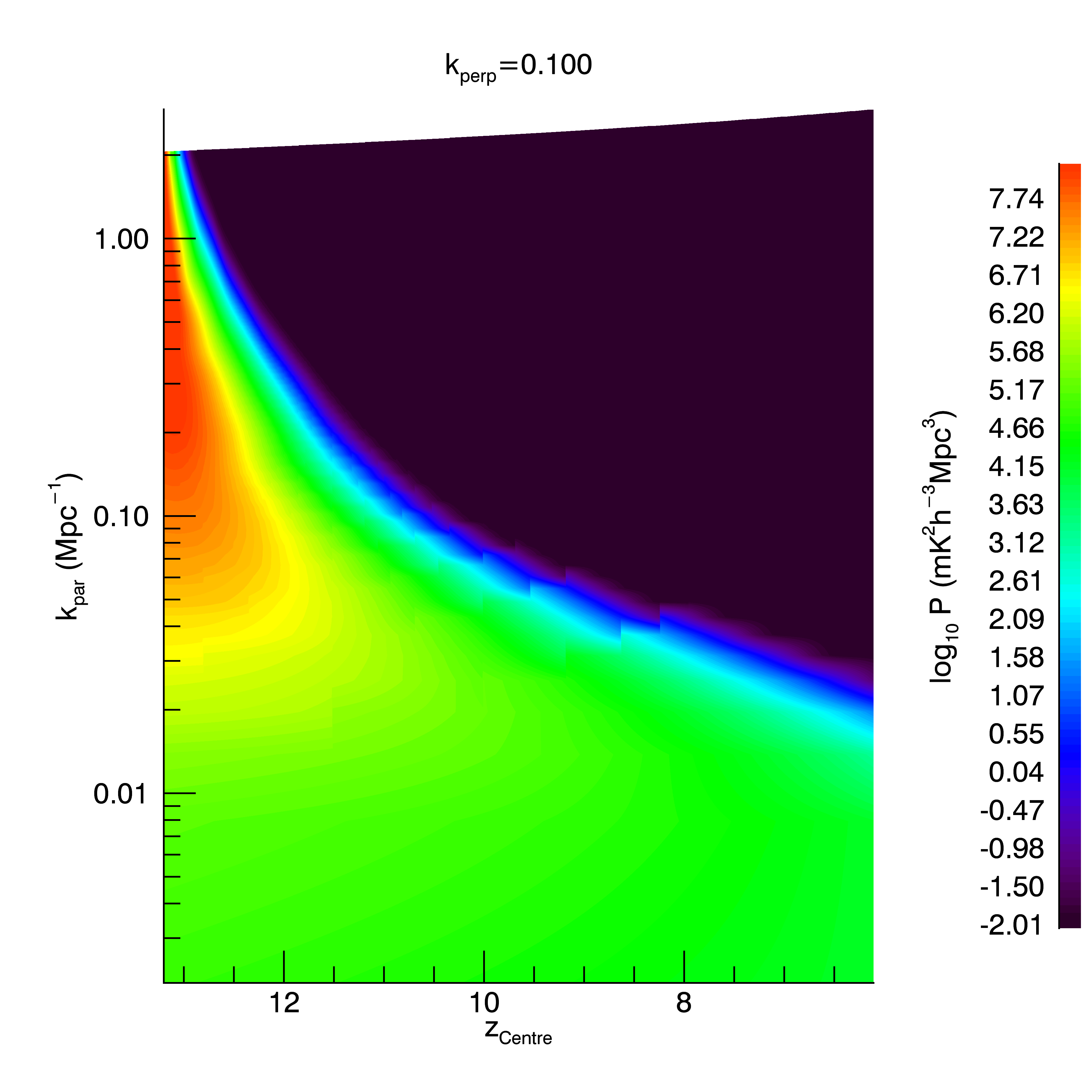}
}
\subfloat[Ratio of power: $P_{0.001}/P_{0.010}$.]{
\includegraphics[width=.45\textwidth]{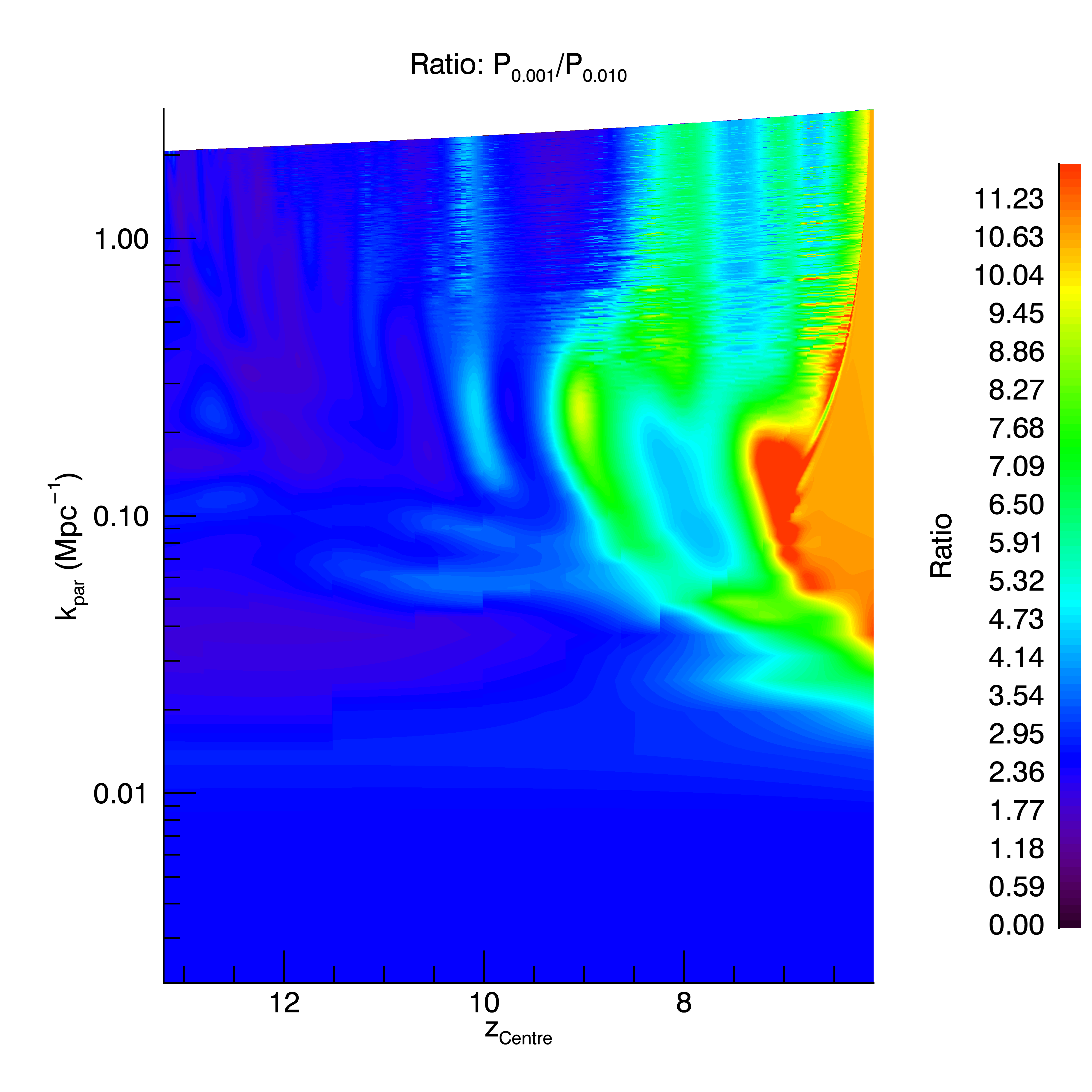}
}
\caption{Power (mK$^2{h}^{-3}$~Mpc$^3$) in Toy Model 1 computed through the Morlet Transform in three angular modes, and the ratio of power, $P_{0.001}/P_{0.010}$.}
\label{fig:gs_ps}
\end{figure*}

We begin by computing the power in Toy Model 1 of 21~cm brightness temperature fluctuations using the Morlet Wavelet Transform, at three different angular scales, $k_\bot$. Figure \ref{fig:gs_ps} displays three scales of interest to EoR and Cosmic Dawn studies, $k_\bot=$~0.001, 0.010, 0.100 $h$~Mpc$^{-1}$, and the ratio of two of these scales.
The signal behaves as expected for the model --- higher redshifts have larger signal, due to the larger brightness temperature fluctuation; larger angular scales are prominent at lower redshifts. The ratio shows the prominence of more large scale power at low redshift.

The parameter estimation precision results are displayed in Table \ref{table:results} for Toy Model 1 and Table \ref{table:toymodel2} for Toy Model 2. The latter contains fewer entries because many of the Fourier Transform estimators over small bandwidth do not have any information to constrain the absorption redshift. It also is computed at a different spatial wavenumber, and so the precision is not comparable between Toy Models 1 and 2.
\begin{table}
	\centering
	\caption{Toy Model 1: parameters and their estimation precision for two experiment total times (T=100~h, 500~h; $k$~=~0.1~$h^{-1}$Mpc).}
	\label{table:results}
	\begin{tabular}{l|cc} 
		\hline
		{\bf T=100~h} & & \\\hline
		$A_s$~(mK) & & \\\hline
		Experiment & Value & Precision (1$\sigma$)\\
		\hline
		100~MHz (MWT) & 30.8~mK & 14.2~mK\\
		8~MHz ($z=6.6$) & 30.8~mK & 7$\times$10$^9$~mK\\
		8~MHz ($z=8.6$) & 30.8~mK & 5$\times$10$^6$~mK\\
		8~MHz ($z=10.6$) & 30.8~mK & 3$\times$10$^3$~mK\\
		8~MHz ($z=12.6$) & 30.8~mK & 20.4~mK\\
		8~MHz (full band) & 30.8~mK & 20.2~mK\\
		\hline\hline
		$\sigma_{\rm ref}$~($h^{-1}$~Mpc) & & \\\hline
		Experiment & Value & Precision (1$\sigma$)\\
		\hline
		100~MHz (MWT) & 100~$h^{-1}$~Mpc & 35.9~$h^{-1}$~Mpc\\
		8~MHz ($z=6.6$) & 100~$h^{-1}$~Mpc & 2$\times$10$^9$~$h^{-1}$~Mpc\\
		8~MHz ($z=8.6$) & 100~$h^{-1}$~Mpc & 1$\times$10$^6$~$h^{-1}$~Mpc\\
		8~MHz ($z=10.6$) & 100~$h^{-1}$~Mpc & 2$\times$10$^3$~$h^{-1}$~Mpc\\
		8~MHz ($z=12.6$) & 100~$h^{-1}$~Mpc & 61.9~$h^{-1}$~Mpc\\
		8~MHz (full band) & 100~$h^{-1}$~Mpc & 59.3~$h^{-1}$~Mpc\\
		\hline\hline
		{\bf T=500~h} & & \\\hline
		$A_s$~(mK) & & \\\hline
		Experiment & Value & Precision (1$\sigma$)\\
		\hline
		100~MHz (MWT) & 30.8~mK & 7.6~mK\\
		8~MHz ($z=6.6$) & 30.8~mK & 1$\times$10$^9$~mK\\
		8~MHz ($z=8.6$) & 30.8~mK & 5$\times$10$^5$~mK\\
		8~MHz ($z=10.6$) & 30.8~mK & 650~mK\\
		8~MHz ($z=12.6$) & 30.8~mK & 10.4~mK\\
		8~MHz (full band) & 30.8~mK & 10.3~mK\\
		\hline\hline
		$\sigma_{\rm ref}$~($h^{-1}$~Mpc) & & \\\hline
		Experiment & Value & Precision (1$\sigma$)\\
		\hline
		100~MHz (MWT) & 100~$h^{-1}$~Mpc & 17.6~$h^{-1}$~Mpc\\
		8~MHz ($z=6.6$) & 100~$h^{-1}$~Mpc & 5$\times$10$^8$~$h^{-1}$~Mpc\\
		8~MHz ($z=8.6$) & 100~$h^{-1}$~Mpc & 2$\times$10$^5$~$h^{-1}$~Mpc\\
		8~MHz ($z=10.6$) & 100~$h^{-1}$~Mpc & 539~$h^{-1}$~Mpc\\
		8~MHz ($z=12.6$) & 100~$h^{-1}$~Mpc & 29.6~$h^{-1}$~Mpc\\
		8~MHz (full band) & 100~$h^{-1}$~Mpc & 27.4~$h^{-1}$~Mpc\\
		\hline
	\end{tabular}
\end{table}
\begin{table}
	\centering
	\caption{Toy Model 2: parameters and their estimation precision for two experiment total times (T=100~h, 500~h; $k$~=~0.01~$h^{-1}$Mpc).}
	\label{table:toymodel2}
	\begin{tabular}{l|cc} 
		\hline
		{\bf T=100~h} & & \\\hline
		$A_s$~(mK) & & \\\hline
		Experiment & Value & Precision (1$\sigma$)\\
		\hline
		100~MHz (MWT) & 30.8~mK & 5.7~mK\\
		8~MHz ($z=10.6$) & 30.8~mK & 20.8~mK\\
		\hline\hline
		$\sigma_{\rm ref}$~($h^{-1}$~Mpc) & & \\\hline
		Experiment & Value & Precision (1$\sigma$)\\
		\hline
		100~MHz (MWT) & 100~$h^{-1}$~Mpc & 31.6~$h^{-1}$~Mpc\\
		8~MHz ($z=10.6$) & 100~$h^{-1}$~Mpc & 164.5~$h^{-1}$~Mpc\\
		\hline\hline
		$z_{\rm dip}$ &  & \\\hline
		Experiment & Value & Precision (1$\sigma$)\\\hline
		100~MHz (MWT) & 10 & 0.02\\
		8~MHz ($z=10.6$) & 10 & 0.17\\
		\hline\hline
		{\bf T=500~h} & & \\\hline
		$A_s$~(mK) & & \\\hline
		Experiment & Value & Precision (1$\sigma$)\\
		\hline
		100~MHz (MWT) & 30.8~mK & 4.4~mK\\
		8~MHz ($z=10.6$) & 30.8~mK & 6.1~mK\\
		\hline\hline
		$\sigma_{\rm ref}$~($h^{-1}$~Mpc) & & \\\hline
		Experiment & Value & Precision (1$\sigma$)\\
		\hline
		100~MHz (MWT) & 100~$h^{-1}$~Mpc & 25.4~$h^{-1}$~Mpc\\
		8~MHz ($z=10.6$) & 100~$h^{-1}$~Mpc & 49.3~$h^{-1}$~Mpc\\
		\hline\hline
		$z_{\rm dip}$ &  & \\\hline
		Experiment & Value & Precision (1$\sigma$)\\\hline
		100~MHz (MWT) & 10 & 0.01\\
		8~MHz ($z=10.6$) & 10 & 0.05\\
		\hline
	\end{tabular}
\end{table}
Use of the full bandwidth as one estimate with the Morlet Transform (`100~MHz (MWT)') yields improved estimation performance, compared with the smaller, individual bands (`8~MHz'). The domination of foregrounds and thermal noise at low redshift, compared with the weakening 21~cm signal, yields poor performance, while higher redshifts are more usable. Toy Model 2 only yields estimable parameters for the full bandwidth, and for a smaller bandwidth that encompasses the absorption feature. For other bandwidths, the absorption redshift cannot be estimated. \textit{This demonstrates the usefulness of using the full bandwidth when the signal structure is unknown.}

Even by combining contiguous 8~MHz bands with a weighted sum in Toy Model 1 to form a full bandwidth estimate with a boxcar sampling of individual 8~MHz bands, (`8~MHz (full band)'), the estimation performance does not improve significantly, due to the relatively very limited information at lower redshifts, and the additional information available by considering all of the data in one estimate.

Note that these results are appropriate for the particular, simple toy models we have employed here. However, the general conclusions are that increased bandwidth provides additional information, which translates to improved estimation performance, \textit{even in the presence of increased noise and bright foregrounds at high redshift, and that local parameters with unknown redshift can be estimated using the full bandwidth transform.}

\section{Discussion and conclusions}
Precise and unbiased parameter estimation are crucial for discriminating different physical models of structure and ionization field evolution in the early Universe. The Morlet Transform, using a large instantaneous bandwidth has the following advantages over Fourier Transforms over fixed bands:
\begin{enumerate}
\item Precision -- use of the full bandwidth in a single estimator incorporates the maximal information available, whereas individual or combined smaller bandwidths yield poorer performance;
\item Bias -- the dynamic scaling of the Morlet envelope to match the line-of-sight wavenumber naturally allows for information of different scales to be probed, whereas Fourier Transforms over wide bandwidths have the potential to dilute and bias real signal evolution. Given that the input signal is \textit{a priori} unknown, this feature is crucial for accurate parameter estimation and model discrimination;
\item Foregrounds -- foregrounds are contained in low wavenumbers in the MWT, as they are in the FFT. However, the containment is better for the MWT because the full lever arm of the large instantaneous bandwidth is available to sample the smooth foreground structure. With smaller bandwidths of the FFT, the foregrounds naturally leak power into higher $k_\parallel$ modes due to the fewer sampled frequencies (spectral leakage).
\end{enumerate}

These elements, taken together, demonstrate the potential power of wavelets (or other localised spectral transform) to achieve improved results compared with the traditional Fourier Transform, and will become more relevant for upcoming large bandwidth experiments, such as that with the SKA.

The estimation analysis presented here only considers the amount of information available to a given transform, in light of the light cone effect. If information is destroyed in the transform (particularly through the squaring process, where loss of phase information impacts some parameter estimation), then the ideal estimation precision will be reduced. In general, both the Fourier Transform and Morlet Transform retain phase information, but formation of the power spectrum, which is common to both, destroys it. In principle, armed with a full understanding of the transform being undertaken, information is not lost and the analysis presented here is sufficient.

A further caveat is the simplicity of the toy models used here. The advantage of Toy Model 1 is its analytic expression and differentiability, but the disadvantage is that it does not reproduce reality. However, it also does not fully exploit the potential benefits of the Morlet Transform. This is because the reionisation parameters being estimated are global, and not local to a particular redshift range. In a more realistic and complex reionisation history, some parameters of interest are highly non-linear in redshift, and a local analysis, such as is provided by the Morlet Transform, will be highly advantageous. This is the key to accessing full local information, as exists due to the light cone effect. Future work will apply the analysis to realistic reionisation simulations \citep[such as 21cmFAST,][]{mesinger07,mesinger11}, where finite differencing, or MCMC, can be used to quantify the estimation precision for key parameters. In addition, parameter bias can be assessed by applying real estimators to the data.

Toy Model 2 aims to address this by including a localised feature. In that case, the benefits of the Morlet Transform become apparent, if the underlying redshift of the absorption feature is unknown.

In this work we have shown an initial demonstration of a localised wavelet transform to explore the Epoch of Reionisation and Cosmic Dawn. It has the potential to yield unbiased information within the context of the light cone effect, without having to resort to loosely-defined finite bandwidths for observation. With the current suite of wide bandwidth EoR experiments, and the future potential 300~MHz instantaneous bandwidth of the SKA, such methods that fully utilise the information will be crucial for estimating parameters of interest and discriminating reionisation history models.

\section*{Acknowledgements}
This research was supported under the Australian Research Council's Discovery Early Career Researcher funding scheme (project number DE140100316), and the  Centre for All-sky Astrophysics (an Australian Research Council Centre of Excellence funded by grant CE110001020). This work was supported by resources provided by the Pawsey Supercomputing Centre with funding from the Australian Government and the Government of Western Australia. We acknowledge the iVEC Petabyte Data Store, the Initiative in Innovative Computing and the CUDA Center for Excellence sponsored by NVIDIA at Harvard University, and the International Centre for Radio Astronomy Research (ICRAR), a Joint Venture of Curtin University and The University of Western Australia, funded by the Western Australian State government.




\bibliographystyle{mnras}
\bibliography{pubs} 



\bsp	
\label{lastpage}
\end{document}